\documentclass[journal, onecolumn]{IEEEtran}

\usepackage{graphicx}      
\usepackage{natbib}        
\usepackage{gensymb}
\usepackage{amsfonts}
\usepackage{amsmath}
\usepackage{bm}
\usepackage[lined,algo2e]{algorithm2e}
\usepackage[utf8]{inputenc} 

\newtheorem{rem}{Remark}
\DeclareUnicodeCharacter{FB01}{fi}
\begin{document}
\title{LiDAR-Based Navigation of Tethered Drone Formations in an Unknown Environment\footnote{This is an extended version of a paper with the same name, presented at IFAC2020.}
} 
\author{M. Bolognini, L. Fagiano%
\thanks{This research has been supported by the Italian Ministry of University and Research (MIUR) under the PRIN 2017 grant n. 201732RS94 ``Systems of Tethered Multicopters''. Corresponding author: M. Bolognini}%
\thanks{Dipartimento di Elettronica, Informazione e Bioingegneria, Politecnico di Milano, Piazza Leonardo da Vinci 32, Milano, Italy. e-mail: michele.bolognini@polimi.it, lorenzo.fagiano@polimi.it}}%
\maketitle

\begin{abstract}                
The problem of navigating a formation of interconnected tethered drones, named STEM (System of TEthered Multicopters), in an unknown environment is considered. The tethers feed electrical power from a ground station to the drones and also serve as communication links. The presence of more than one interconnected drone provides enough degrees of freedom to navigate in a cluttered area. The leader drone in the formation must reach a given point of interest, while the followers must move accordingly, avoiding interference with the obstacles. The challenges are the uncertainty in the environment, with obstacles of unknown shape and position,  the use of LiDAR (Light Detection And Ranging) sensors,  providing only partial information of the surroundings of each drone, and the presence of the tethers, which must not impact with the obstacles and pose additional constraints to how the drones can move. To cope with these problems, a novel real-time planning algorithm based on numerical optimization is proposed: the reference position of each drone is chosen in a centralized way via a convex program, where the LiDAR scans are used to approximate the free space and the drones are moved towards suitably defined intermediate goals in order to eventually reach the point of interest. The approach is successfully tested in numerical simulations with a realistic model of the system.
\end{abstract}



\section{Introduction}
The drone market has been rapidly expanding in the last decade, and it is expected to keep growing in the coming years. Technological advancements and decreasing prices are driving the interest in the use of unmanned aerial vehicles  (UAVs) for various purposes, such as mapping (\cite{mappingCoralReefs}), emergency response (\cite{postEarthquakeResponse}) and building inspection (\cite{UASReviewBuiltEnvironment}). Tethered drones are employed in applications where long operational time is needed and/or the system must be mechanically anchored for safety. Different aspects of tethered drones have been investigated (\cite{Tognon2016,Lee2015,Naldi2012,Lupashin2013,Oh2006,Choi2014}) and several commercial products exist as well (see, e.g., \cite{Elistair2016}).\\
Differently from the mentioned contributions, where a single tethered drone is considered, our research explores a novel tethered multi-drone system, called STEM (System of TEthered Multicopters), proposed in \cite{STEMIFAC}. STEM consists of a formation of two or more multicopter drones, tethered to each other and to a ground station. 
The tethers transfer electrical power from a ground power source to the drones, and also provide a power line communication network. Their length can be adjusted automatically by onboard winches, according to the formation's geometry. Each drone is stabilized by a local feedback controller, while a high level, centralized control algorithm computes the reference positions for all the units, according to the specific mission. The use of more than one tethered drone is justified by the gained flexibility: a chain of drones can reach places, e.g. behind obstacles, that a single tethered drone can not. A prototype STEM featuring two drones, each with 12 kg of mass and connected by 100-m-long cables, is currently under development at Politecnico di Milano.\\
In this paper, we study in particular the problem of autonomous navigation of STEM in a three-dimensional (3D) environment, with a-priori unknown obstacles.
The goal is to reach a given point of interest with the first drone of the series. The other drones in the formation have to adapt their trajectories to allow the first one to reach its goal, while making all the units (and the tethers) avoid the obstacles. Since the latter are supposed to be a priori unknown, the centralized navigation algorithm can rely only on the partial information gathered by each unit, including, in addition to the usual position and attitude of each drone, the readings of planar LiDAR (Light Detection And Ranging) sensors installed on the UAVs.\\
This problem can be classified as a collision avoidance one, for which the methods exist in literature ranging from \textit{planning} techniques, such as graph search (\cite{GraphSearch, OctreeAStar,DuBK14}), RRT (\cite{Li2016}), and potential fields (\cite{PotentialFieldsGlobal}), to \textit{reactive} techniques, e.g., \cite{DVZReactive}, \cite{1DInsectFly}. However, to the best of the authors' knowledge, none of the existing approaches can deal with the combination of aspects in the problem at hand, i.e.: the presence of the tethers, 
the fact that the position and shape of the obstacles is assumed to be fully unknown a priori and the use of LiDAR sensors, each one returning the distance of the closest obstacles along a finite number of directions.\\ 
The main contribution of this paper is a new reactive approach to deal with these aspects. A real-time algorithm collects the information from the drones and employs it to build suitable polytopic constraints. The latter approximate locally the set of all directions that do not impact with an obstacle, while also ensuring that any two subsequent drones remain in line-of-sight. The presence of tether catenary is accounted for by a constraint tightening procedure. Then, a quadratic program (QP) is solved to update the reference position tracked by each drone, in order to eventually reach the point of interest.  To formulate the approach and test it in a realistic simulation environment, additional contributions are presented, including a model of the LiDAR sensors and a model of the tether as a nonlinear multi-body dynamical system.

\section{System model and problem formulation}
\label{sec:description}
\begin{figure}
\begin{center}
\includegraphics[trim={11cm 3.5cm 8cm 4.5cm},clip,width=0.4\columnwidth, height=4cm]{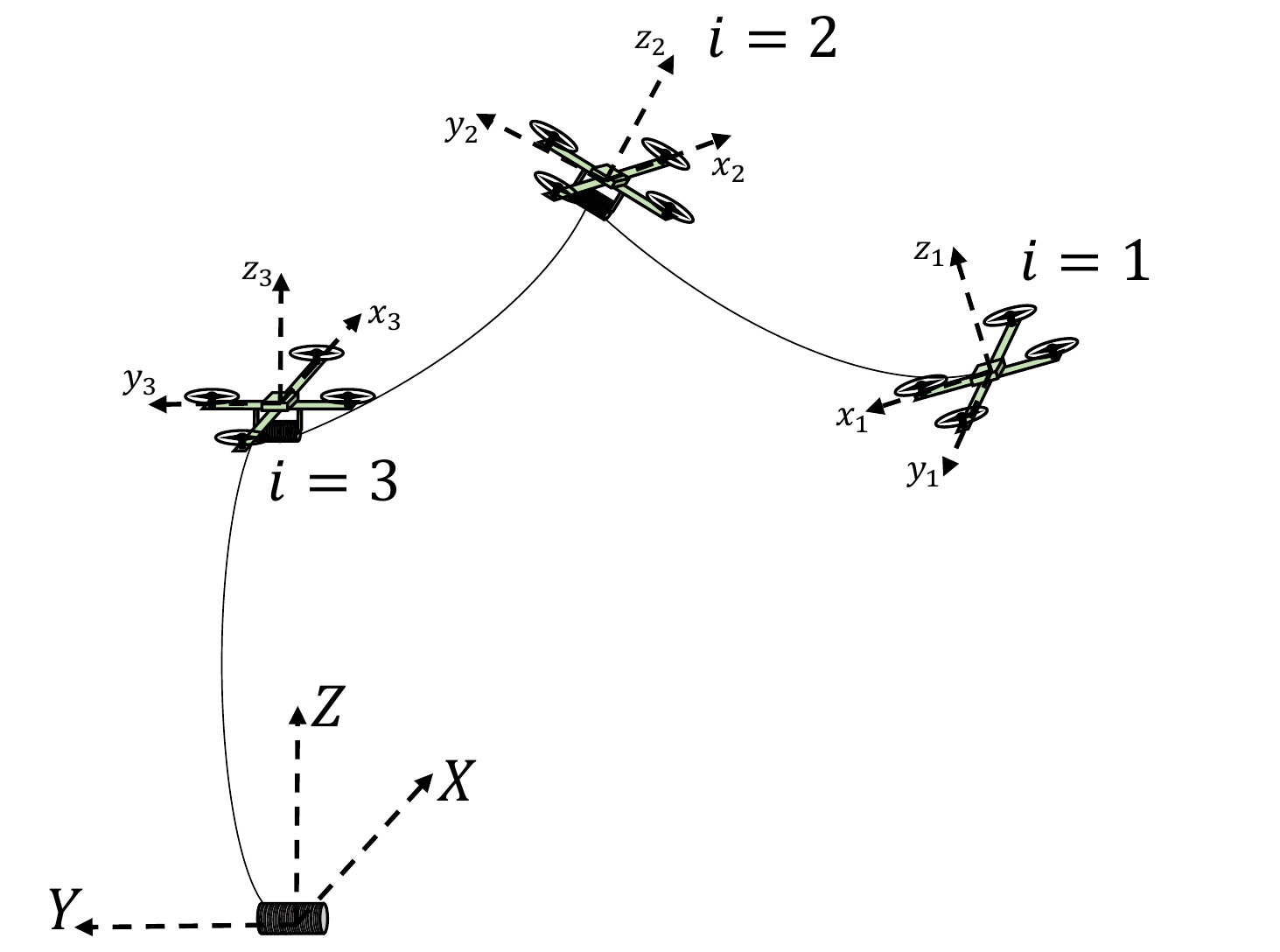}    
\caption{Model of a considered STEM system with three drones.} 
\label{fig:sys_sketch}
\end{center}
\end{figure}
The navigation algorithm we propose in this paper operates in discrete time with a chosen sampling period $T_s$. At each sampling time, it collects information from the drones and computes and sends back reference position values. Since it 
does not employ past information nor it features an internal state, for notational simplicity we drop the explicit time-dependence of all the involved variables.
\subsection{Multicopter drones}\label{ss:drones}
To simulate the system, we employ six-degrees-of-freedom, nonlinear continuous-time models of the drones, as described in \cite{STEMIFAC}. The considered setup features a ground station and $N \in \mathbb{N}$ drones in series, identified by indexes $i = 1, 2, ..., N$.  The drone with  $i=1$, named leader, is the farthest one in the series from the ground station, see Fig. \ref{fig:sys_sketch}. The position and velocity vectors of the $i$-th drone in a fixed, right-handed inertial reference frame ($X,Y,Z$), with $Z$ pointing up, are denoted with $\bm{p}_i, \dot{\bm{p}}_i\in\mathbb{R}^3$, where $\bm{p}_i=[p_{X,i},p_{Y,i},p_{Z,i}]^T$ and  $\cdot^T$ is the matrix transpose operation. The origin of the reference frame coincides with the position of the ground station.  Electro-actuated winches are installed on the ground station and on each drone, except the leader, and can adjust the length of the tethers connecting any two subsequent units in the series.  We denote with $L_i$ the length of the reeled-out tether connecting drone $i$ to drone $i+1$ (and drone $N$ to the ground station). The value of $L_i$ is thus available from the local controller of the winch on drone $i+1$ (or ground station). The drones and the winches are assumed to be automatically controlled with local feedback loops that make the UAVs track a reference position vector, denoted with $\bm{p}_{ref,i}$, while automatically adjusting the tether lengths. A possible design approach for such local controllers is provided in \cite{STEMIFAC}.
\subsection{LiDAR sensors}\label{SS:lidars}
\begin{figure}[!hbt]
\begin{center}
\includegraphics[width=\columnwidth]{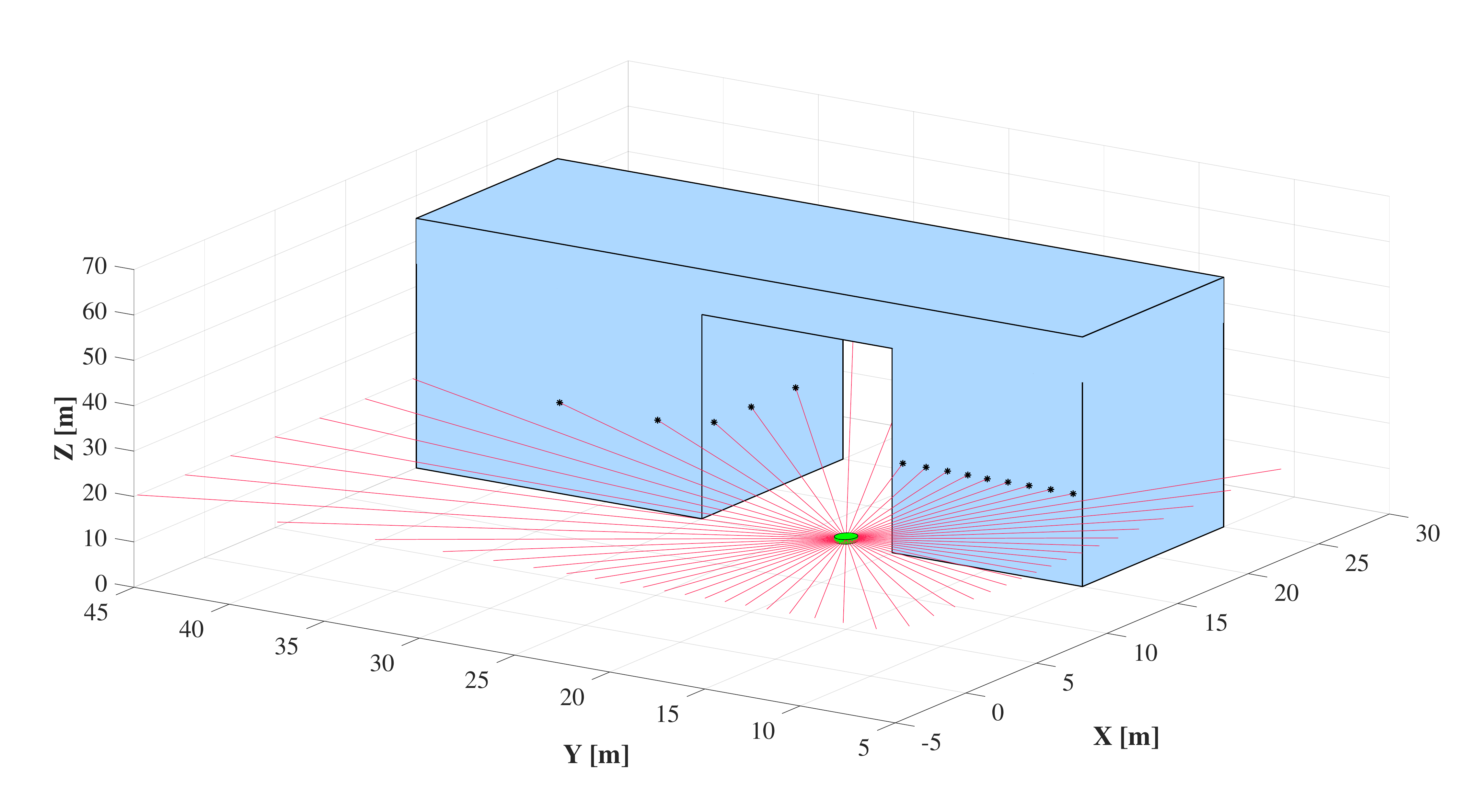}    
\caption{Representation of a LiDAR scan along the horizontal plane in a simulation environment. The sensor is mounted on a UAV (Unmanned Aerial Vehicle), represented as a green circle, while the light blue overarching structure represents an obstacle. Each red line emanating from the vehicle is a laser beam, while black dots represent the position estimates of obstacles detected by the LiDAR.} 
\label{fig:LidarRepresentation}
\end{center}
\end{figure}

The LiDAR sensors are modeled as point-like masses, as their shape and appearance do not influence their simulation.
A planar sensor can have an arbitrary angular span $\Theta$, defined as the angle between the first and last laser beams it shoots.
Without loss of generality, we assume $\Theta = 360 \degree$, \textit{i.e.} the sensor scans an entire plane and has no blind spots.
Not all real sensors have a $360 \degree$ span, but more of them can be installed on the same platform to obtain an equivalent effect.
Secondly, LiDAR sensors are also characterized by an angular resolution $\theta_s$, defined as the angle between two consecutive laser beams.
A sensor scan thus produces a vector $\boldsymbol{s} \in \mathbb{R}^{M}$ with $M = \lfloor\frac{\Theta}{\theta_s}\rfloor$ entries $s(j)$, with $ j = 0,\dots,M-1$, each one corresponding to the distance of the closest obstacle along the  direction $j \theta_s$ (with $j=0$ corresponding to a selected, fixed  direction). For example in the horizontal plane $XY$, if we assume that the LiDAR sensor is centered in the origin and $j=0$ corresponds to the $X-$axis, we can describe the measured obstacles positions in Cartesian coordinates as:
\begin{equation}\label{eq:osbtacles_coord}
\begin{array}{rcl}
X_{j}^{obs} &=& s_i(j) \cos(j \theta_s)\\
Y_{j}^{obs} &= &s_i(j) \sin(j \theta_s); \quad j = 0,\ldots, M-1,
\end{array}
\end{equation}
see Fig. \ref{fig:LidarRepresentation} for a visualization.

In simulation, the vector $\boldsymbol{s}$ is computed via a collision detection routine between the obstacles, defined by a series of possibly overlapping polytopes, and $M$ segments originating from the drones and directed along the probed directions.
The LiDARs also have a maximum detection range $s_{max}$ (30$\,$m for the LiDARs considered here): if in a given direction $j \theta_s$ no obstacle is detected within this range, then $s(j)=\infty$.
From a practical standpoint, once a direction is selected, each point along that direction, starting from the sensor position up until the maximum radius has been reached, is checked to determine if it is inside or outside an obstacle.
Each consecutive point is taken at a distance $d$ from the previous one, where $d$ is a tunable parameter. The smaller $d$ is, the more precise the simulation becomes, at the cost of greater computational burden.
Given the approximation introduced by sampling with step $d$, it is useful to note that the maximum error is $e_{max} = \frac{d}{2}$.
It is also possible to obtain a more realistic simulation by adding an arbitrary Gaussian noise to the estimate.
Each drone $i$ is equipped with two planar LiDAR sensors, spanning $360$ degrees each in one of two perpendicular planes. We assume the LiDARs to be installed on active gimbals that stabilize their attitude notwithstanding the drone's motion. Referring to Fig. \ref{fig:sys_sketch}, one plane, named ``horizontal'', is always parallel to $(X,Y)$ and elevated at the drone's $Z-$coordinate,  while the other one, the ``vertical plane'', is perpendicular to  $(X,Y)$ and can be oriented by adjusting the gimbal's yaw angle. This allows each drone to detect obstacles lying inside one of such planes, while no information is gathered outside them.\\
The choice of LiDARs for navigation is advantageous because they are fast, hence it is possible to use them in real-time scenarios, they are active sensors, thus they work in conditions where light and contrast are scarce or excessive, and they retain accuracy at a distance. 

\subsection{Tethers}\label{SS:tethers}
Accounting for the tether catenary is crucial to develop and test a sensible navigation algorithm in presence of obstacles.
Here, we consider a multi-body approach 
, in which the $i-$th tether is  modeled as a chain of $N_t$ inner nodes with mass $m_{t,i}$, computed as:
\begin{equation}\label{eq:tether_parameters}
\begin{array}{rcl}
m_{t,i} & = & \dfrac{L_i\rho_t}{N_t}, 
\end{array}
\end{equation}
where $\rho_t$ is the tether mass per unit of length. Each mode has associated position and velocity vectors, $\bm{p}_{t,i,l}, \dot{\bm{p}}_{t,i,l}\in\mathbb{R}^3,l=1,\ldots,N_t$. By adding the points at the extremes, which are assumed to be fixed to the devices attached to the tether (i.e., a drone or the ground station),  we thus have $N_t+1$ segments, and each inner node is subject to its own weight and to the forces applied by the two neighboring tether segments. We consider only elastic and internal friction forces, and neglect the aerodynamic ones, assuming that little wind is present and the drones' speed values are relatively small.
The tether segments are modeled as nonlinear springs, which can transfer forces only when the tether segment is taut, in parallel with linear dampers. 
For the $l$-th node, the equations of motion thus read:
\begin{equation}\label{eq:node_i_motion}
\ddot{\bm{p}}_{t,i,l} =\dfrac{\left(\bm{F}_{e,i,l+1}+\bm{F}_{e,i,l-1}+\bm{F}_{f,i,l+1}+\bm{F}_{f,i,l-1}\right)}{m_t}-\left[
\begin{array}{c}
0\\0\\g
\end{array}\right]
\end{equation}
where $g$ is the gravity acceleration, $\boldsymbol{F}_f$ represent friction forces and $\boldsymbol{F}_e$ elastic ones. For the first and last nodes, i.e., $l=1$ and $l=N_t$, the position and velocity vectors of the tether ends are considered in the computation of tether forces. These are equal to the position and velocity vectors of the devices attached to the tether ends, i.e., drone $i$ and $i+1$, which in turn are subject to the force vectors related to the first and last segment, with a minus sign, thus coupling the tether model with those of the connected systems. The time-varying, nonlinear dynamical system that describes the tethers 
is used in our simulation tests (see Section \ref{S:results}), and also to pre-compute a map which, on the basis of the relative position of the two tether extremes and on the tether length, returns the relative height $\underline{Z}_i$ of the lowest point of the tether's catenary, assumed at steady-state 
:
\begin{equation}\label{eq:lookup_table}
\underline{Z}_i=f(\bm{p}_i,\bm{p}_{i+1},L_i)
\end{equation}

\subsection{Problem Formulation}
\begin{figure}[b]
	\begin{center}
		\includegraphics[width=.5\columnwidth]{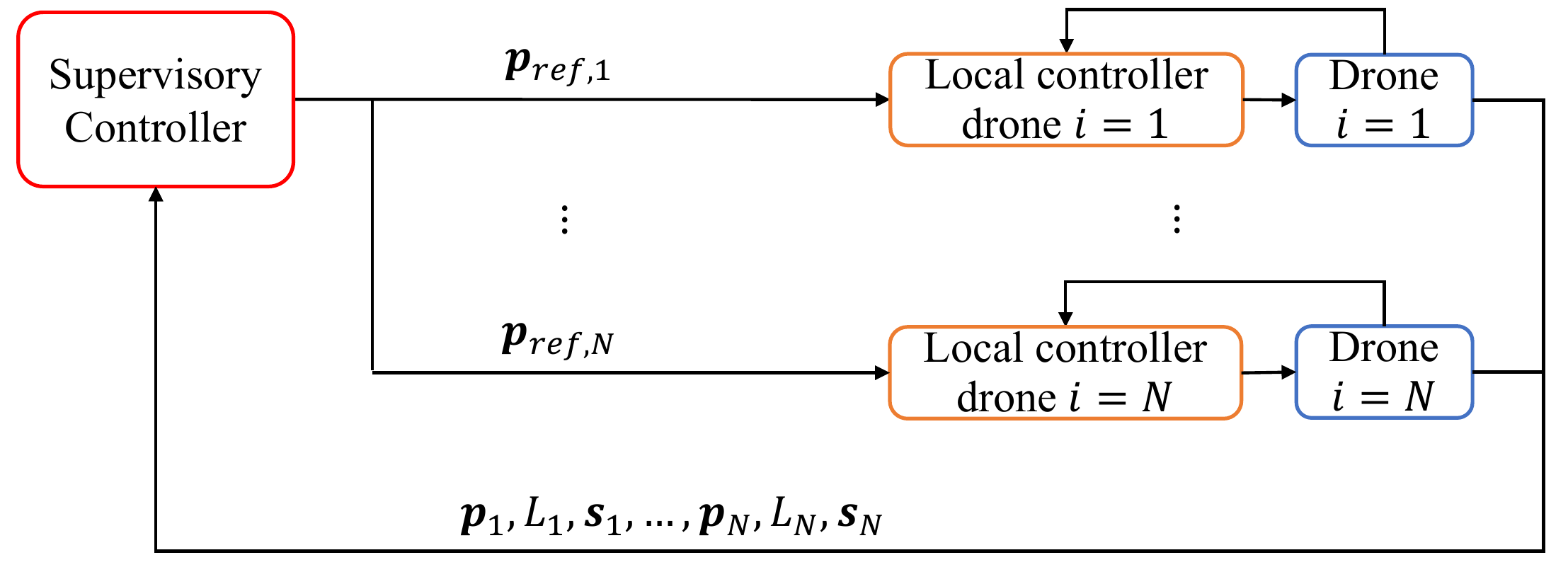}    
		\caption{Layout of the considered control system. At each time step, the supervisory controller calculates position reference values for all drones, based on their current positions and LiDAR sensor readings. } 
		\label{fig:ControlScheme}
	\end{center}
\end{figure}
The main focus of this work is the supervisory controller highlighted in Fig. \ref{fig:ControlScheme}. 
Its inputs are the position vectors $\bm{p}_i,\,i=1,\ldots,N$, the tether lengths $L_i,\,i=1,\ldots,N$, and the LiDAR readings, $\bm{s}_i,\,i=1,\ldots,N$ sent by the drones, while the outputs are the reference position vectors $\bm{p}_{ref,i},\,i=1,\ldots,N$  to be computed and sent to the drones in each sampling period. The problem we address is how to design such a supervisory controller to guide the leader drone to a point of interest with coordinates $\bm{p}_{poi}$ in the inertial reference $(X,Y,Z)$, specified by an external agent, while moving the overall formation in order to avoid collisions between the drones and/or the tethers and the obstacles.	

\section{Real-time navigation and obstacle-avoidance algorithm for STEM}\label{S:solution}
\begin{figure}[!hbt]
	\centering
	\includegraphics[width=8.4cm]{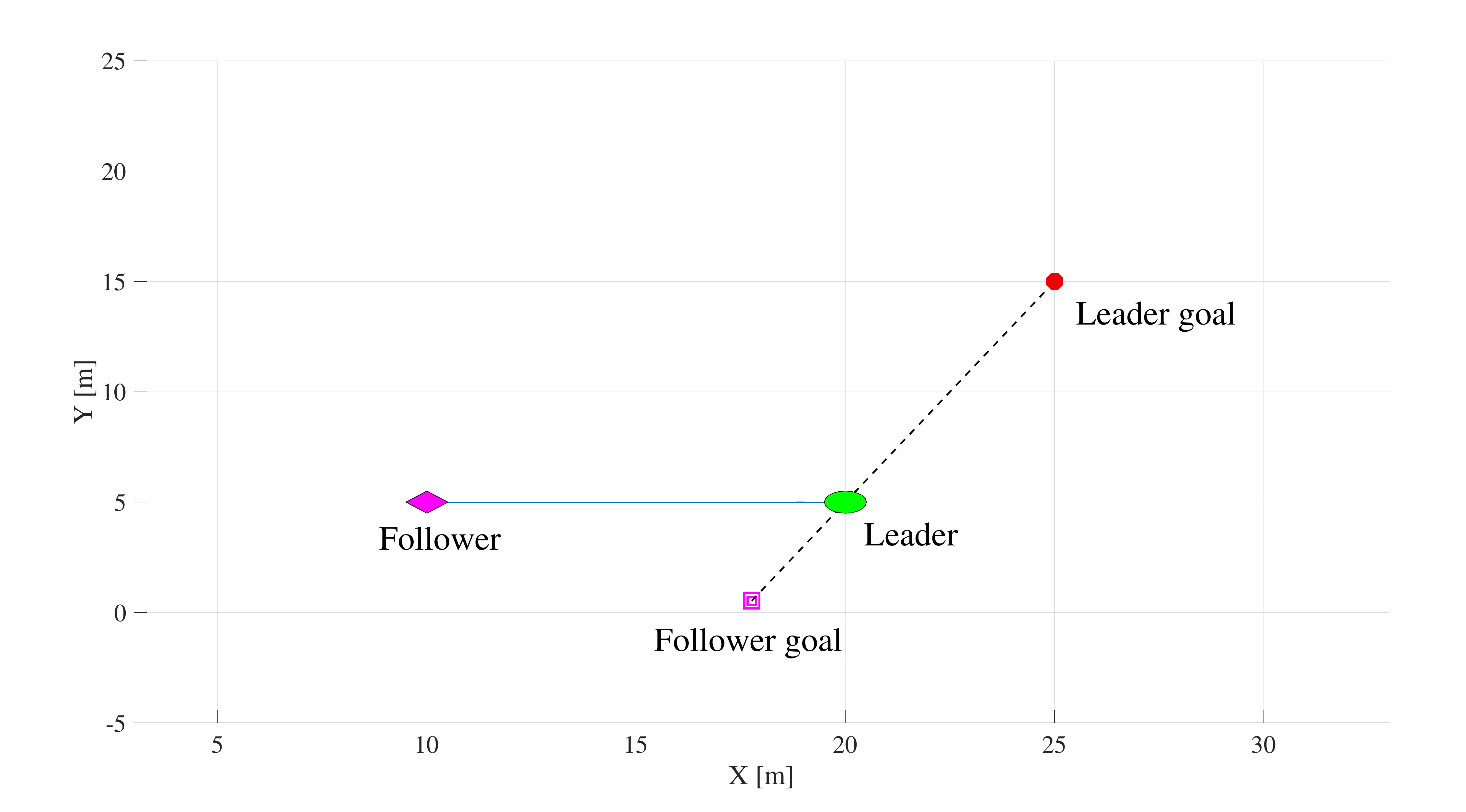}
	\caption [Goal calculation for follower drone]{\label{fig:FollowerSetPoint} Computation of goals on a plane parallel to $(X,Y)$. The goal of drone $i+1$ is computed as a point lying at a user-defined distance $\bar{d}$ (here $5$ m) behind drone $i$ on the line connecting $\bm{p}_{goal,i}$ and $\bm{p}_i$.}
\end{figure}

Let us collect the decision variables in a single vector $\boldsymbol{x} = [\bm{p}_{ref,1}^T,\ldots, \bm{p}_{ref,N}^T]^T\in\mathbb{R}^{3N}$, and let us define the matrix $\bar{\Lambda}=\text{diag}(\Lambda_1,\ldots,\Lambda_N)\in\mathbb{R}^{3N\times3N}$, where $\text{diag}(\cdot)$ denotes a block-diagonal matrix and, for $i=1,\ldots,N$, $\Lambda_i$ are $3\times3$ diagonal matrices with the tuning parameters $\lambda_i\in(0,1)$ on the diagonal. The proposed navigation algorithm consists of the following main steps:
\begin{enumerate}
	\item Collect the information $\bm{p}_i,\,L_i$, and $\bm{s}_i,\,i=1,\ldots,N$;
	\item Using the collected information, formulate suitable \emph{goals} for the drones, expressed as position vectors $\bm{p}_{goal,i}\in \mathbb{R}^{3}$ in the inertial frame. Collect the goals in vector $\bm{x}_{goal} = [\bm{p}_{goal,1}^T,\ldots, \bm{p}_{goal,N}^T]^T$;
	\item Also based on collected information, build  a matrix $A\in\mathbb{R}^{p\times3N}$ and vector $\bm{b}\in\mathbb{R}^p$ that define polytopic constraints on the drones' reference positions;
	\item Compute $\bm{x}^*$ as:
	\begin{equation}\label{eq:algo_QP}
	\begin{array}{rc}
	\bm{x}^*=&\arg\min\limits_{\bm{x}} (\bm{x}-\bm{x}_{goal})^T\bar{\Lambda}(\bm{x}-\bm{x}_{goal})\\
	&\text{subject to}\\ 
	&A\bm{x} \leq \bm{b}
	\end{array}
	\end{equation}
	\item Send the optimal position references contained in $\bm{x}^*$ to the corresponding drones, repeat from (1) at the next time step.
\end{enumerate}
Note that \eqref{eq:algo_QP} is a strictly convex QP of small size with Hessian $\bar{\Lambda}$, a diagonal matrix, which can be solved in very short time, see e.g. \cite{WaBo08,Bempo16}, and is thus suitable for real-time application. The weights $\gamma_i$ are used to trade off, in the cost function, the distance of the different drones drone from their goals. Typically, a larger weight $\gamma_1$ is assigned to the leader's cost with respect to the other ones.
The key points of the approach are the computation of the goals $\bm{p}_{goal,i},i=1,\ldots,N$ and of $A$ and $\bm{b}$.\\ 
Regarding the goals, we initialize $\bm{p}_{1}^{goal}=\bm{p}_{poi}$ (i.e., the leader's goal is always equal to the wanted point of interest), and proceed sequentially, by setting the goal 
of drone $i+1$ as:
\begin{equation}\label{eq:goal_comp}
\bm{p}_{goal,i+1} = \bm{p}_{goal,i} - \bar{d} \frac{\bm{p}_{goal,i}-\bm{p}_i}{\left\lVert  \bm{p}_{goal,i}-\bm{p}_i\right\lVert},
\end{equation}
where $\bar{d}$ is a design parameter and $\bm{p}_i$ the current position of drone $i$. Equation \eqref{eq:goal_comp} corresponds to the geometric construction presented in Fig. \ref{fig:FollowerSetPoint}, whose aim is to make each drone $i+1$ point a target  ``behind'' drone $i$ on the line connecting the latter to its goal, at a distance $\bar{d}$.
Regarding the constraints in \eqref{eq:algo_QP}, for simplicity we first consider a 2D case, and later extend the approach to 3D.
\subsection{Computation of constraints in 2D}\label{SS:goals_constr_2D}
For the 2D case, let us assume that all drones have the same $Z$ coordinate, so that their movements are bound to a plane parallel to $(X,Y)$. We start by considering each  drone $i$ individually. Recall that the LiDAR readings $\bm{s}_i$ feature a finite entry if an obstacle is detected along the corresponding direction within the maximum distance, and infinite otherwise. It is then safe to state that if the drone moves along one of the directions with infinite distance in $\bm{s}_i$, no collisions can occur until the next sampling period. Moreover, if the scans are dense enough (\textit{i.e.} if $\theta_s$ is small), then it is reasonable to assume that no extremely thin obstacle lies between two consecutive scans, at least at a close distance. Figure \ref{fig:LidarRepresentation} shows a graphical interpretation of this concept. Note that the maximum distance introduced in Section \ref{SS:lidars} can also be set to a value smaller than the LiDAR's limit, for example to navigate in narrow environments surrounded by obstacles, and that the intended sampling period (e.g., 0.1$\,$s)  is small compared to the drones' speed (e.g., 2 m/s).
With these considerations in mind, let us define the following indexes:
\begin{equation}
\label{eq:j_limits}
\begin{array}{rccll}
\underline{j}_i&=&\min\limits_{j=1,\ldots,M} &j &:s_i(j)=\infty\\
\bar{j}_i&=&\max\limits_{j=\underline{j}_i+1,\ldots,M} &j& :s_i(k)=\infty,k=\underline{j}_i+1,\ldots,j
\end{array}
\end{equation}
the pair $(\underline{j}_i,\bar{j}_i)$ thus identifies an angular sector spanning directions with no visible obstacles. The two lines that limit such a sector are given by:
\begin{align}
\label{eqn:firstConstraintLine}
\underbrace{\begin{bmatrix}
-\tan(\underline{j}_i\theta_s) & 1\\
-\tan(\bar{j}_i\theta_s) & 1
\end{bmatrix}}\limits_{A_i\left(\underline{j}_i,\bar{j}_i\right)}
\begin{bmatrix}
p_X\\p_Y
\end{bmatrix} = 
\underbrace{\begin{bmatrix}
-\tan(\underline{j}_i\theta_s)p_{X,i}+p_{Y,i}\\
-\tan(\bar{j}_i\theta_s)p_{X,i}+p_{Y,i}
\end{bmatrix}}\limits_{\bm{b}_i\left(\underline{j}_i,\bar{j}_i\right)}
\end{align}
Now, assuming that $\theta_s\left(\bar{j}_i-\underline{j}_i\right)\leq\pi$, the following inequality constraints (possibly with a change of sign to select the half-planes where no obstacles are present) identify the obstacle-free sector  as a polyhedron:
\begin{equation}
\label{eq:angular_sector}
A_i\begin{bmatrix}
p_X\\p_Y
\end{bmatrix}\leq \bm{b}_i
\end{equation}

\begin{figure}[]
\centering
\includegraphics[width=8.4cm]{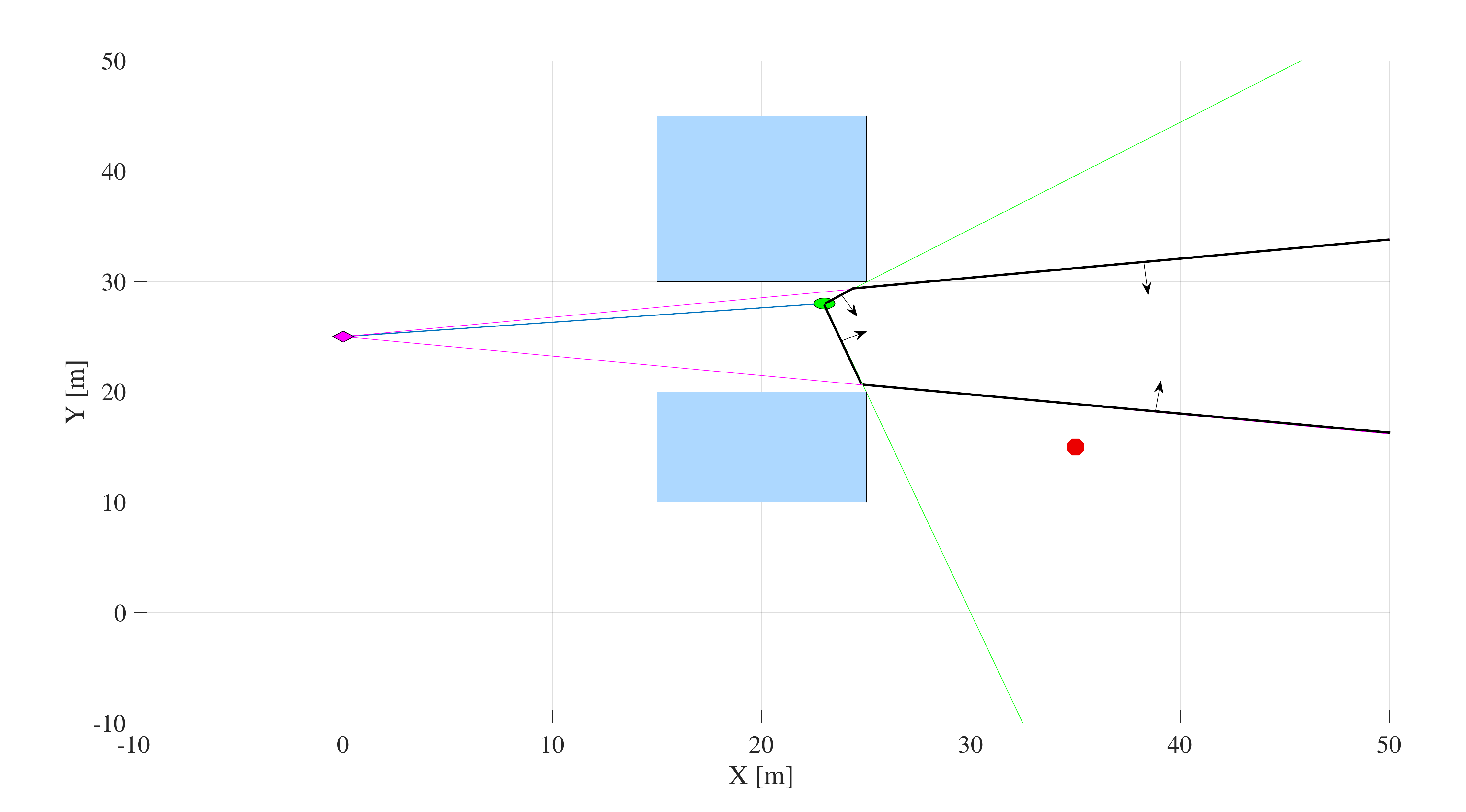}
\caption [Influence of follower's constraints on leader]{\label{fig:LFConstraints}The feasible space for the position reference of drone $i$ (green circle) is limited by the one seen by drone $i+1$ (magenta square). If drone $i$ were to travel within its own constraints only, the tether might impact with the obstacles.}
\end{figure}

\begin{rem}\label{rem:constraints}The following considerations are due:
	\begin{enumerate}
\item If $\underline{j}_i\theta_s$ and/or $\bar{j}_i\theta_s$ are equal to $\dfrac{\pi}{2}$ or $\dfrac{3\pi}{2}$, then the corresponding row of matrix $A_i\left(\underline{j}_i,\bar{j}_i\right)$ and vector $\bm{b}_i\left(\underline{j}_i,\bar{j}_i\right)$, become, respectively, $[1\; 0]$ and $p_{X,i}$;
\item If $N_c>1$ non-contiguous, obstacle-free angular sectors are present, then a corresponding number of pairs $\left(\underline{j}_{i,l},\bar{j}_{i,l}\right),\,l=1,\ldots,N_c,$ is extracted from vector $\bm{s}_i$ by minor modification of \eqref{eq:j_limits}. Our algorithm eventually needs one of these pairs to be selected. This is done by the following approach:
	\begin{equation}
	\label{eq:constraints_selection}
	A_i=A_i\left(\underline{j}_{i,l^*},\bar{j}_{i,l^*}\right),\;\,\bm{b}_i=\bm{b}_i\left(\underline{j}_{i,l^*},\bar{j}_{i,l^*}\right)
	\end{equation}
where
\begin{equation}\label{eq:constr_select_procedure}
	l^*=\arg\max\limits_{l=1,\ldots,N_c}c_l
\end{equation}	
	and
\begin{equation}\label{eq:constr_select_procedure2}
	\begin{array}{c}
	c_l=\left\|\left[\begin{array}{cc}
	\cos{\left(\underline{j}_{i,l}\theta_s\right)}&\sin{\left(\underline{j}_{i,l}\theta_s\right)}\\
	\cos{\left(\bar{j}_{i,l}\theta_s\right)}&\sin{\left(\bar{j}_{i,l}\theta_s\right)}
	\end{array}
	\right]\left[\begin{array}{c}
	(p_{X,goal,i}-p_{X,i})\\
	(p_{Y,goal,i}-p_{Y,i})
	\end{array}\right]\right\|_\infty
	\end{array}
\end{equation}\normalsize
Namely, \eqref{eq:constraints_selection}-\eqref{eq:constr_select_procedure2} select the angular sector whose borders feature the largest inner product with the vector pointing from the drone to its goal. This angular sector is deemed the most promising one on the basis of the currently available information;
		\item If $\theta_s\left(\bar{j}_i-\underline{j}_i\right)>\pi$, then the sector spanning the obstacle-free directions does not correspond to a convex set. In this case, with a procedure similar to \eqref{eq:constraints_selection}-\eqref{eq:constr_select_procedure2}, the line with smallest absolute inner product with vector $(\bm{p}_{goal,i}-\bm{p}_i)$ is dropped, thus turning the angular sector into a half-plane. 
		\item If the point of interest lies between the leader drone and an obstacle, the most obvious movement, that brings the leader directly to its goal in a straight line, is prevented by the presence of the obstacle behind the goal. This situation is detectable by comparing the distance between drone 1 and its goal with the values in $\bm{s}_1$ in the relevant directions. If this situation is detected, the sensor readings can be suitably manipulated before the calculation of constraints takes place, in order to enable movement in the angular sector where the point of interest lies.
\end{enumerate}
\end{rem}
%
Once $A_i$, $\bm{b}_i,\,i=1,\ldots,N$ have been computed for all drones, we can build the following constraints on the reference position for each drone:
\begin{equation}
\label{eq:constraints_single_drone}
\bar{A}_i\begin{bmatrix}
p_{X,ref,i}\\p_{Y,ref,i}
\end{bmatrix}\leq \bar{\bm{b}}_i
\end{equation}
where $\bar{A}_N=A_i$, $\bar{\bm{b}}_N=\bm{b}_N$ and, for $i<N$:
\begin{equation}
\label{eq:constraints_single_drone2}
\begin{array}{cc}
\bar{A}_i=\text{diag}(A_{i},A_{i+1}),&\bar{\bm{b}}_i=\left[
\begin{array}{c}
\bm{b}_i\\
\bm{b}_{i+1}
\end{array}
\right].
\end{array}
\end{equation}
Namely, for each drone we constrain the reference position to lie inside the intersection of the obstacle-free regions of the drone itself and of the next one in the series (i.e. its follower), in order to keep the two UAVs within line-of-sight of each other, thus ensuring that the $i-$th tether between them does not impact with any obstacle, see Fig. \ref{fig:LFConstraints} for a visualization. 
This approach also favors recursive feasibility of the optimization problem at each time step,  because drone $i+1$ moves inside an obstacle-free area and drone $i$ moves inside a subset of it. However, in some situations the presence of new, previously unseen obstacles might render constraints \eqref{eq:constraints_single_drone} unfeasible. In this case, one can either reduce the distance above which a scanned direction is deemed obstacle-free, or consider a different obstacle free-angular sector (if more than one are present) for one of the two drones. As a matter of fact, in our tests recursive feasibility was always present. A formal proof of this property is subject of ongoing research.\\
The final step to build $A$ and $\bm{b}$ in \eqref{eq:algo_QP} is to consider constraints \eqref{eq:constraints_single_drone} altogether:
\begin{equation}
\label{eq:constraints_overall}
\begin{array}{cc}
A=\text{diag}(\bar{A}_{1},\ldots,\bar{A}_{N}),&\bm{b}=\left[
\begin{array}{c}
\bar{\bm{b}}_1\\\vdots\\
\bar{\bm{b}}_N
\end{array}
\right].
\end{array}
\end{equation}

\subsection{Computation of constraints in 3D}
The main difficulty in extending the approach to the third dimension is the fact that each drone features two planar sensors, which is not equivalent to one 3D sensor. Thus,  the points where obstacles are detected  lie in the geometrical union of the two planes where the LiDAR scans take place, and it is not safe to make assumptions on the shape of the obstacles outside such region. For this reason, at each time step the algorithm sets the position reference for each drone in one of these planes and nowhere else. Therefore it must choose, for each drone, along which plane the movement will happen for the current sampling period.  As mentioned in Section \ref{SS:lidars}, we assume that the vertical plane where the LiDAR operates can be rotated via a gimbal. In particular, we assume that the leader carries out the vertical scan in the plane that contains the point of interest, while the other drones $i+1,\,i=1,\ldots,N-1,$ carry out the vertical scan in the plane that contains drone $i$. This is a sensible choice because, as a first approximation, if there is little wind and the speed of the drones is not high, the vertical plane containing drones $i$ and $i+1$ also contains the tether $i$ with its catenary.\\
The constraints in the vertical plane are derived with the same approach adopted for the horizontal plane, i.e. equations \eqref{eq:j_limits}-\eqref{eq:constraints_overall}. Let us denote with $\bar{A}_{H,i},\,\bar{\bm{b}}_{H,i}$ the terms in \eqref{eq:constraints_single_drone} pertaining to the horizontal plane, and with $\bar{A}_{V,i},\,\bar{\bm{b}}_{V,i}$ those related to the vertical one. Then, after the constraint sets in both planes have been computed, in order to choose whether to place the reference position of each drone in the horizontal or vertical plane, the following quantities are derived:
\[
c_H=\left\{
\begin{array}{ll}
0&\text{if } \bar{A}_{H,i}\begin{bmatrix}
p_{X,goal,i}\\p_{Y,goal,i}
\end{bmatrix}\leq\bar{\bm{b}}_{H,i}\\
d\left(
\left[
\begin{array}{c}
p_{X,goal,i}\\
p_{Y,goal,i}
\end{array}\right],\mathcal{C}_H
\right)&\text{otherwise}
\end{array}\right.
\]
\[
c_V=\left\{
\begin{array}{ll}
0&\text{if } \bar{A}_{V,i}\begin{bmatrix}
p_{XY,goal,i}\\p_{Z,goal,i}
\end{bmatrix}\leq\bar{\bm{b}}_{V,i}\\
d\left(
\left[
\begin{array}{c}
p_{XY,goal,i}\\
p_{Z,goal,i}
\end{array}\right],\mathcal{C}_V
\right)&\text{otherwise}
\end{array}\right.
\]
where $p_{XY,goal,i}=\left\|
	\begin{array}{c}
	p_{X,goal,i}-p_{X,i}\\
	p_{Y,goal,i}-p_{Y,i}
	\end{array}\right\|$
is the relative position on plane $(X,Y)$ of the goal with respect to the drone (recall that the vertical plane always contains both points), $d(\bm{v},\mathcal{A})$ is the distance between point $\bm{v}$ and set $\mathcal{A}$, finally $\mathcal{C}_H=\left\{\begin{bmatrix}
p_{X}\\p_{Y}
\end{bmatrix}:\bar{A}_{H,i}\begin{bmatrix}
p_{X}\\p_{Y}
\end{bmatrix}\leq\bar{\bm{b}}_{H,i}\right\}$ and $\mathcal{C}_V=\left\{\begin{bmatrix}
p_{XY}\\p_{Z}
\end{bmatrix}:\bar{A}_{V,i}\begin{bmatrix}
p_{XY}\\p_{Z}
\end{bmatrix}\leq\bar{\bm{b}}_{V,i}\right\}$.\\
If $c_H\leq c_V$, the horizontal plane is chosen, otherwise the vertical one. Intuitively, this approach chooses the plane in which the distance between the corresponding constraint set and the point of interest (possibly projected on $(X,Y)$ for the horizontal plane) is smaller. When the goal is behind an obstacle, this strategy effectively chooses to circumvent it by moving in the plane where the required travel length appears to be smaller. Once the plane has been chosen, the corresponding constraints are included in the QP \eqref{eq:algo_QP}, with the addition of suitable linear constraints to force the position reference to lie on that plane. The constraints pertaining to the discarded plane are instead dropped.\\  
Another aspect that must be included in the 3D case is the presence of the tether catenary. To avoid impact between the tether and an obstacle in the vertical plane, the following linear constraints are also included, exploiting the map \eqref{eq:lookup_table}:
\begin{equation}\label{eq:vertical_constraints}
\begin{aligned}
p_{Z,ref,i} &\geq (\overline{Z}_{obs}+ \Delta Z) + (p_{Z,i}-\underline{Z}_i) \\
p_{Z,ref,i+1} &\geq (\overline{Z}_{obs}+ \Delta Z) + (p_{Z,i+1}-\underline{Z}_i),
\end{aligned}
\end{equation}
where $\overline{Z}_{obs}$ is the highest point detected by the vertical LiDAR sensor of drone $i+1$, oriented towards drone $i$, i.e., the height of the highest detected obstacle below the cable connecting drone $i+1$ to $i$, and  $\Delta Z$ is a tunable safety margin. Note that \eqref{eq:vertical_constraints} are enforced on both drones $i$ and $i+1$: this fact might impact feasibility if, for one or both of the drones, the horizontal plane had been chosen. In these cases, if the problem is infeasible, the vertical plane is chosen instead of the horizontal one, and the related constraints are included in the QP, so that both drones are allowed to move in $Z$ direction. This is the only exception to the plane selection approach described above.\\
%
%
As in the 2D case, LiDARs readings are manipulated if the goal lies between the drone and an obstacle. This is even more important in 3D, since the ground is considered an obstacle and if the goal is lower than the leader's position it is difficult to descend without modifying the sensor output.\\
Finally, we include in the optimization program additional linear constraints pertaining to the reference position of drone $N$, $\bm{p}_{ref,N}$, which prevent it from moving too far away in the $(X,Y)$ plane from the ground station. 

\section{Simulation results}\label{S:results}
All simulations were run with Matlab and Simulink. The latter was employed to simulate the continuous-time dynamics of the drones and the tethers and the discrete-time local controllers in between any two sampling instants of the supervisory controller, implemented in Matlab. The QP \eqref{eq:algo_QP} has been solved with Matlab's \verb|quadprog|. We carried out simulation tests with different shapes of obstacles and positions of the point of interest, covering many realistic application scenarios, and report here some of the obtained results. 
\begin{figure}[]
\centering
\includegraphics[width=8.4cm]{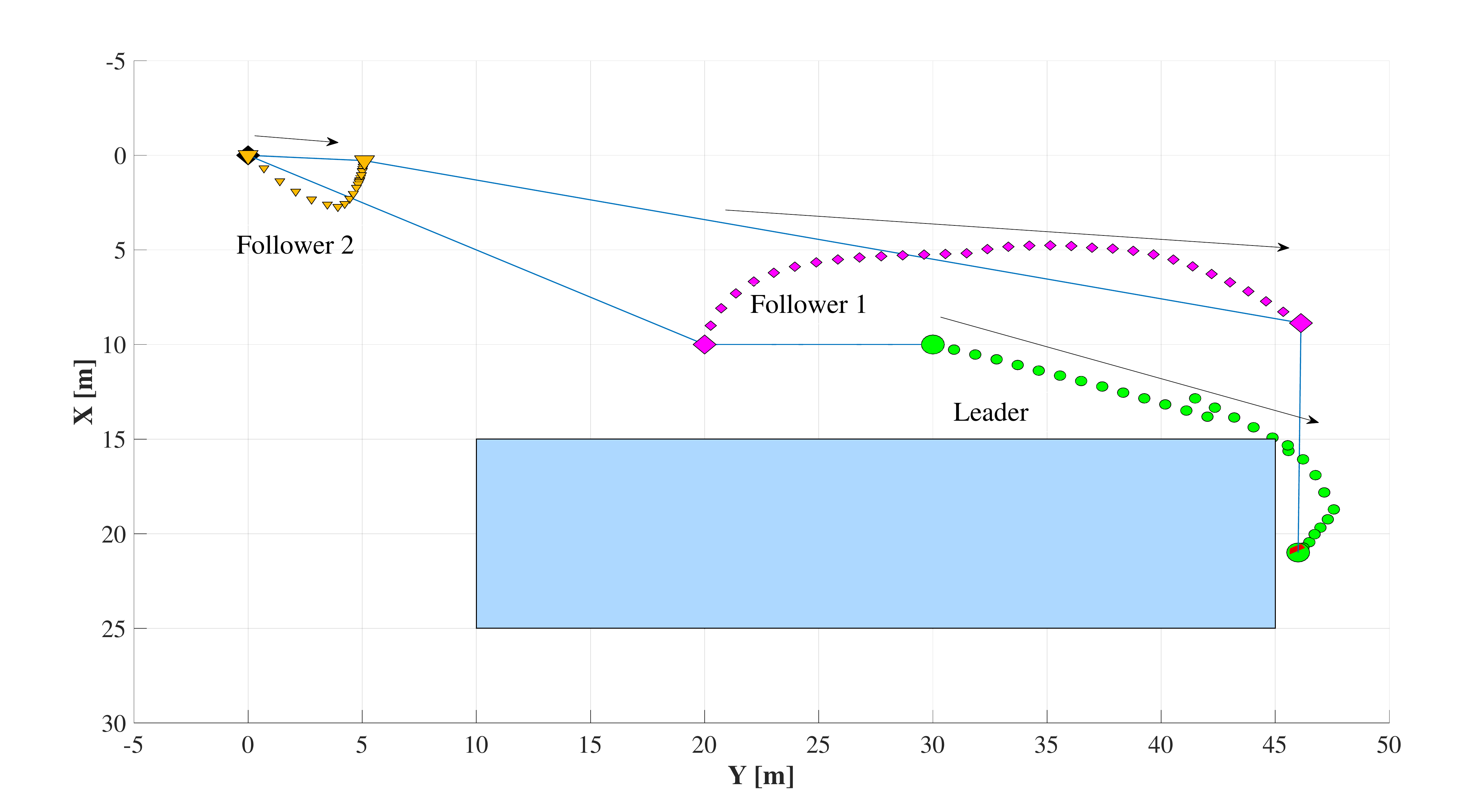}
\includegraphics[width=8.4cm]{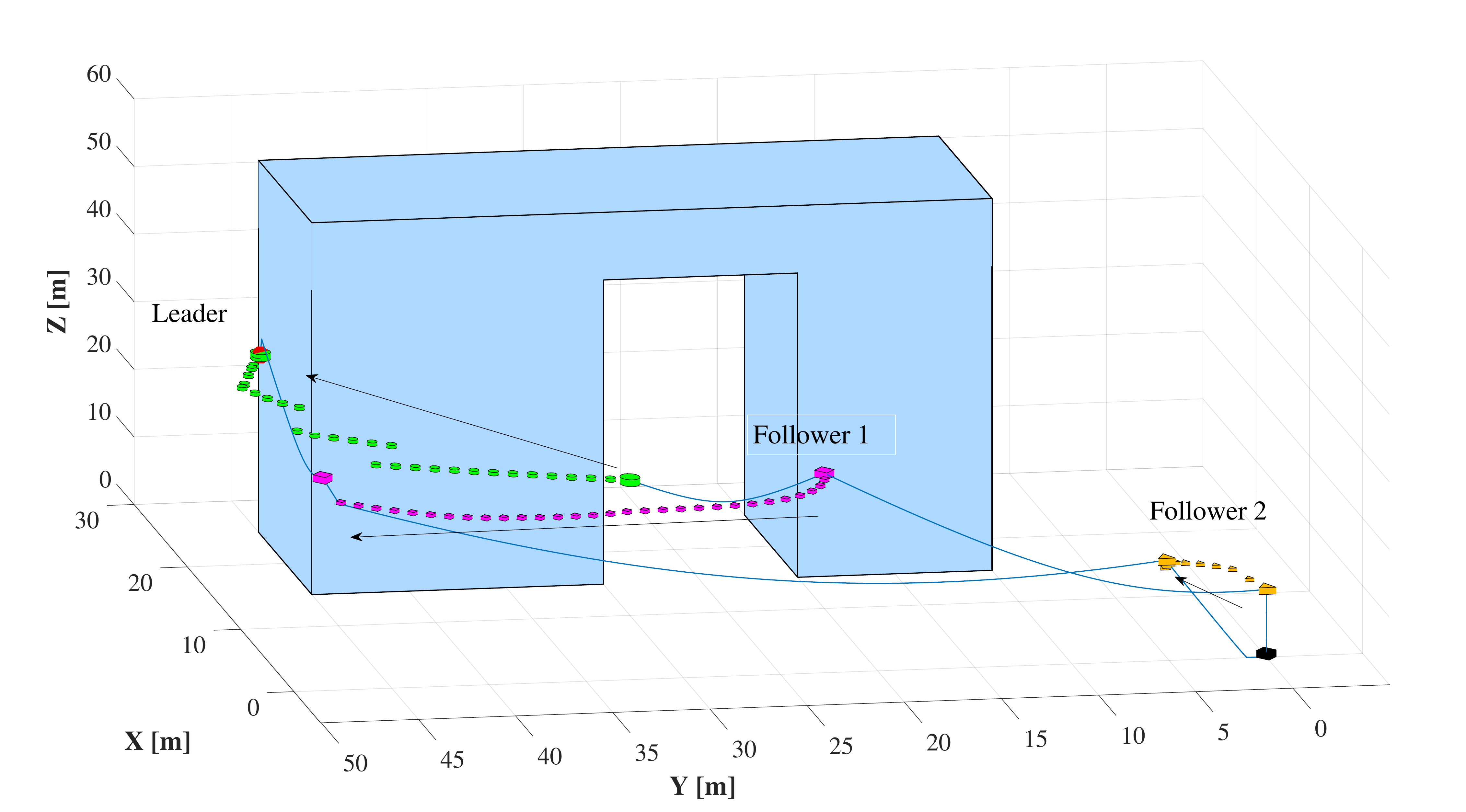}
\caption [Simulation 1]{\label{fig:SimCornerSide}A simulation with only two drones and no ground station (the drones move towards increasing $y$ values). The follower drone limits the feasible space for the leader, temporarily blocking it from reaching the destination, in order to prevent cable collisions.}
\end{figure}
\begin{figure}[]
\centering
\includegraphics[width=8.4cm]{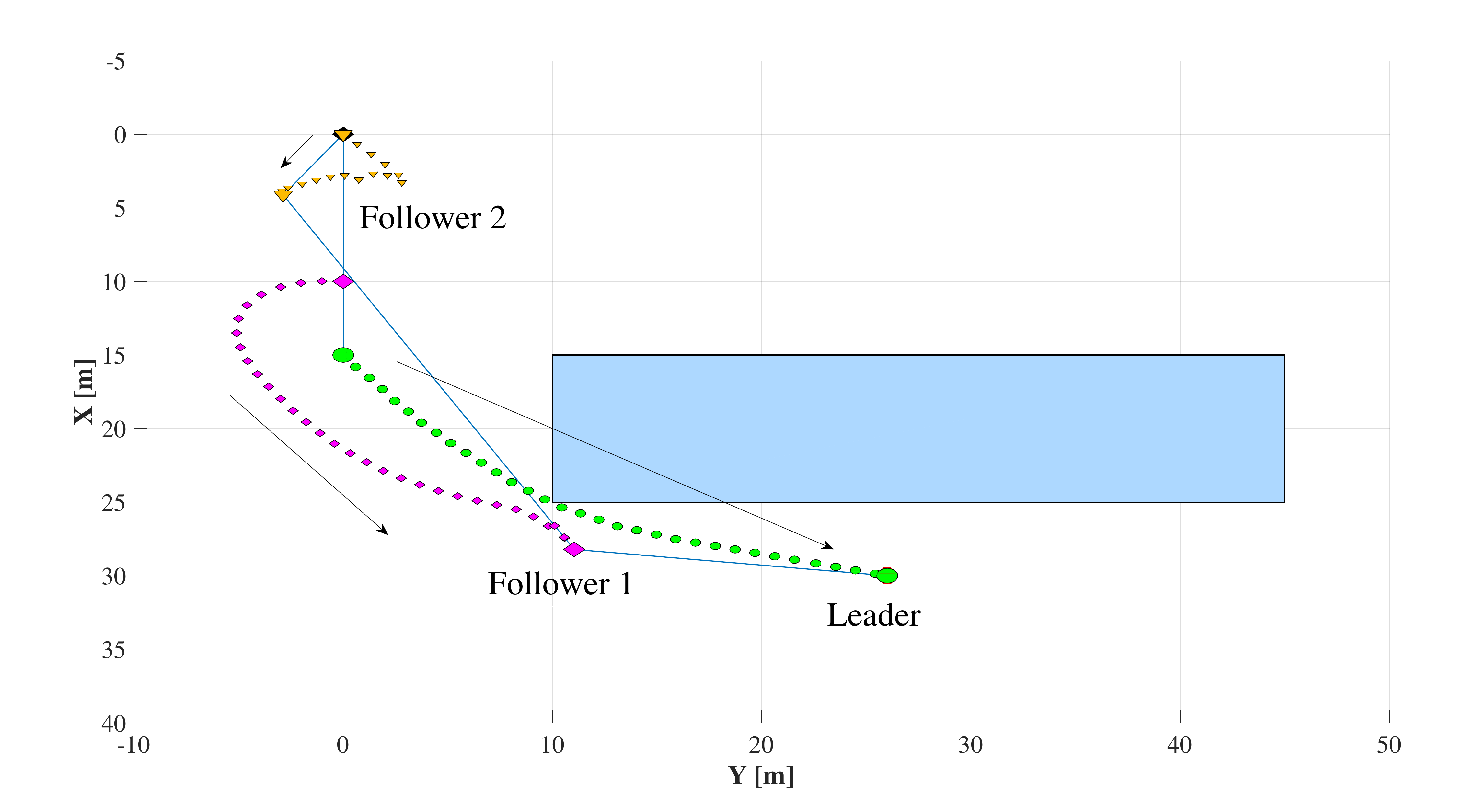}
\includegraphics[width=8.4cm]{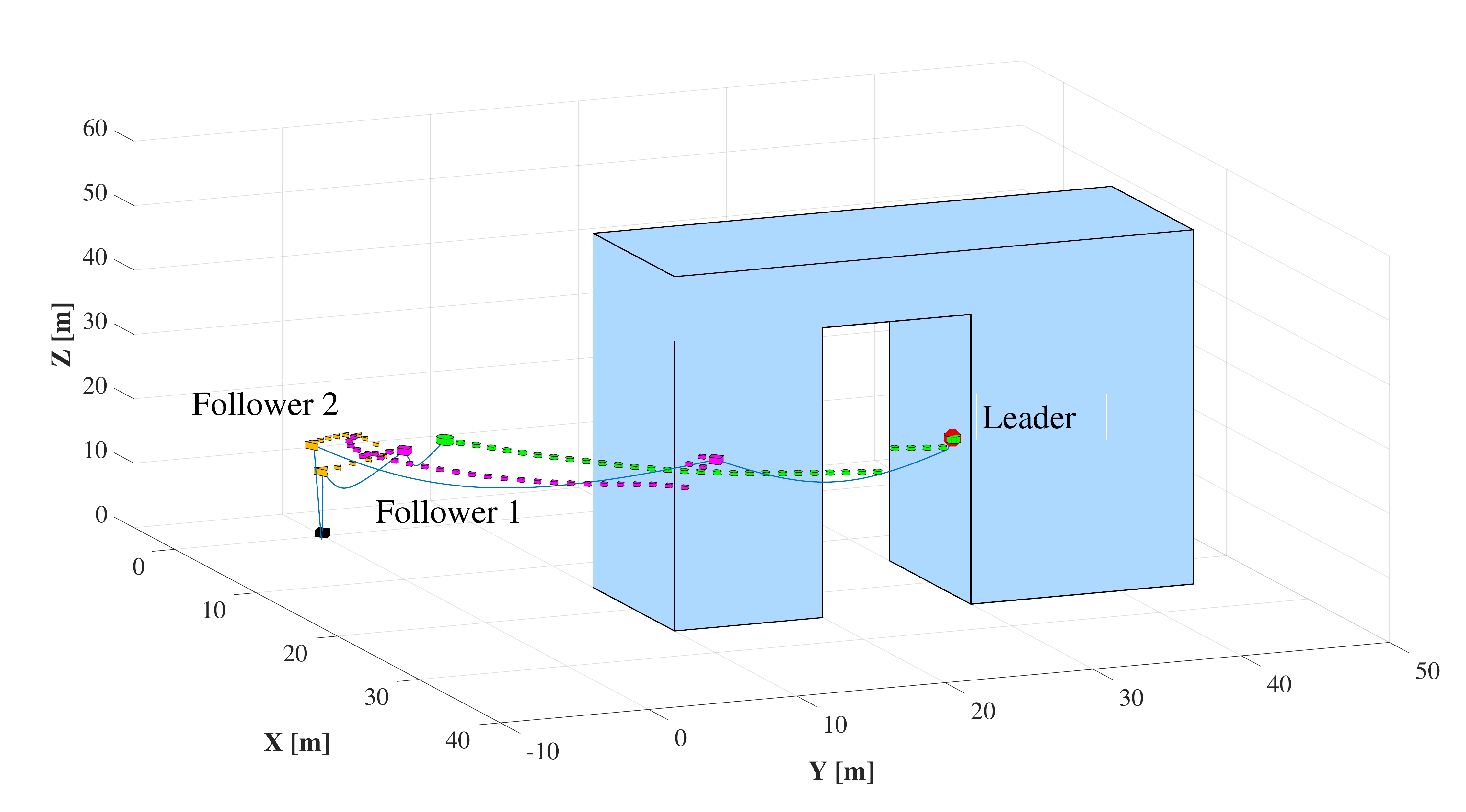}
\caption [Simulation 2]{\label{fig:Sim3DAroundSide}A simulation of the whole system with three drones and a ground station (the drones move towards increasing $x$ and $y$ values). Note how the third drone (yellow triangle) is confined within a restricted area above the ground station.}
\end{figure}
The simulations show that the proposed algorithm can manage a formation of three drones plus the ground station so as to guide the aircrafts in 3D space avoiding the a-priori-unknown obstacles. Figure \ref{fig:SimCornerSide} highlights how the follower drone is guided around an obstacle and the leader is temporarily stopped from approaching the goal by the propagation of constraints introduced in section \ref{SS:goals_constr_2D}, in order to avoid a collision between the cable and the obstacle. The last movements of the leader also show that the condition where the goal lies between  the drone and an obstacle is correctly recognized and taken care of, letting the drone approach the obstacle in order to get to its final destination.
Figure \ref{fig:Sim3DAroundSide} also shows that the leader can reach a goal initially behind an obstacle, furthermore it illustrates other aspects described in this paper:
\begin{itemize}
	\item The third drone,  directly connected to the ground station, hovers close to it in a confined region of space;
	\item The $i-$th drone tends to position itself on the line connecting drone $i-1$ to its goal;
	\item Each drone independently decides whether to move within the horizontal or vertical plane, depending on its goal and surroundings;
	\item The constraints on the horizontal plane are correctly propagated: in fact the second drone, towards the end of the simulation test, remains in a region of space where collision between obstacle and the second cable (i.e., the cable between drones 2 and 3) is avoided, letting its distance from the leader grow.
\end{itemize}
\section{Conclusion}
An algorithm able to navigate a formation of tethered drones in an unknown environment, relying on LiDAR measurements, has been presented. The algorithm operates in real-time using only the current information, and is based on the solution of a convex QP. The algorithm does not employ maps nor it stores information about the visited zones, rather it drives the drone formation to incrementally explore the environment towards the final goal, thus it is suitable for navigation in time-varying environments as well. In the development of this approach, safety was prioritized over performance, because the obstacle-avoidance feature is a primary concern for the automated flight of systems such as STEM. 
This work was aimed at enabling the autonomous flight in a simulation environment of a system of tethered drones, and it represents a first step for further research and development activities. These include:  the inclusion of additional performance indicators such as speed, energy
consumption and repeatability; the addition of mapping and trajectory generation functions; the addition of vision-based sensors such as cameras; the realization of a real-world prototype to experimentally test the different approaches in a building monitoring task.
\bibliographystyle{plainnat}
\bibliography{ifacconf}             

\end{document}